# Wavelet Bicoherence Analysis as a Method for Investigating Coherent Structures in an Electron Beam with an Overcritical Current

**A. A. Koronovskiĭ and A. E. Hramov**

*Saratov State University,*
*Astrakhanskaya ul. 83, Saratov, 410012 Russia*
*e-mail: aeh@cas.ssu.runnet.ru*

**Abstract**—Results are presented from numerical modeling of the effect of the inhomogeneity of the ion background on the complicated spatiotemporal dynamics of an electron beam with a virtual cathode in plane geometry. The possibility is demonstrated of increasing the generation frequency without changing the beam current. The spatiotemporal structures that form in the beam and govern the complicated stochastic dynamics of the nonuniform electron-plasma system under consideration are investigated by the methods of wavelet bicoherence and by analyzing the calculated electron trajectories on the space–time diagrams.

## 1. INTRODUCTION

One of the simplest models of spatially bounded plasma systems is a diode gap penetrated by an electron beam against a charge-neutralizing ion background. Such a model makes it possible to investigate both collective processes in plasmas and the effect of the boundary conditions that give rise to different beam instabilities [1, 2]. In the theory of nonlinear oscillations and waves, interest in spatially bounded plasma systems stems from the possibility of using them to model distributed active media with global couplings.

Here, we consider the model of a plane diode gap. This is a classical model for studying nonradiative instabilities in electron-plasma systems [3, 4]. In 1944, J. Pierce [5] investigated an electrostatic instability of a high-current electron beam completely charge-neutralized by the ion background in a system of finite length. This nonradiative instability, which arises from the finite length of the system, triggers space charge waves whose amplitude grows in time, resulting in the formation of a virtual cathode, which gives rise to a backward electron wave. The conditions for the onset of the instability are determined by the mode composition of the plasma oscillations in the beam. In the potential approximation, the spatial spectrum of the one-dimensional waves of the beam is described by the dispersion relation [2]

$$k_{\parallel}^2 - \frac{\omega_p^2 k_{\parallel}^2}{(\omega - k_{\parallel} v_0)^2} = 0, \qquad (1)$$

where $k_{\parallel}$ is the wavenumber of the beam wave, $\omega_p$ is the plasma frequency, and $v_0$ is the unperturbed beam velocity. With increasing beam current in the system, a frequency range appears in which there exists a backward electron wave, i.e., a wave propagating in the direction opposite to that of the beam. The condition for the generation of a backward wave determines the condition for the onset of an electrostatic instability. A beam with a virtual cathode can be interpreted as an active resonant system [6] in which the initial perturbation of the space charge density is amplified at a frequency of about $\sim 2\omega_p$.

Being simple, however, the system exhibits a wide range of dynamic regimes in the nonlinear stage of electrostatic instability, including the regime of deterministic chaos [7–9]. In some papers (see, e.g., [10–14]), the complicated dynamics of systems with a virtual cathode is attributed to the formation of and the interaction between coherent spatiotemporal structures in an electron beam.

Note that the complicated dynamics of electron-plasma systems, as well as turbulent processes and the formation of structures in them, should be analyzed with allowance for their nonlinearity. To study the latter phenomenon, it is necessary to use special methods for analyzing the formation of coherent structures and their dynamics, because classical linear analytic methods (e.g., spectral analysis) fail to provide sufficient information on the processes occurring in a nonlinear system. In this paper, we consider a technique for applying one of the nonlinear methods of the analysis of complicated processes in distributed systems, namely, bicoherent wavelet transformation, which was first proposed by van Milligen *et al.* [15, 16].

The wavelet analysis, which has been elaborated relatively recently, is a powerful tool for studying the dynamics of nonlinear systems. It is attracting increased attention [17–20] and, in particular, has begun to be used in studies of the complicated nonlinear dynamics of plasmas [21–24]. The wavelet analysis allows one to resolve the structural details of the signal in time and to separate out the scales of the corresponding dynamic processes in space. Also, the wavelet transformation makes it possible to overcome one of the drawbacks of applying the spectral analysis to turbulence and chaos, namely, the expansion of the signal in question in harmonic components. For the nonlinear equations describing the complicated phenomena in distributed systems, this is not the case: they do not possess harmonic eigenmodes.

The bispectra (bicoherence functions) characterize the phase relations (or phase coupling) between different frequency components of the signal spectrum (see, e.g., [25]). We can speak of the phase coupling when the signal to be analyzed contains two components such that the sum (or difference) of their frequencies $\omega_1$ and $\omega_2$ and the sum of their phases $\phi_1$ and $\phi_2$ both remain constant during a certain time interval. The wavelet bicoherence analysis makes it possible to resolve the spatiotemporal data into coherent structures, i.e., to determine the internal beam structure, which is governed by the phase coupling between the coherent structures, and to find out how the beam structure evolves in time. That is why the wavelet bicoherence analysis provides a useful tool for resolving and investigating coherent structures that form in spatially distributed systems.

Here, we present the results from numerical modeling of the effect of the inhomogeneity of the ion background on the complicated spatiotemporal dynamics of an electron beam with an overcritical current in plane geometry. We propose a method in which the wavelet bicoherence analysis is used to resolve the spatiotemporal data into coherent structures. We then apply this method to investigate the spatiotemporal structures that form in a beam and govern the complicated stochastic dynamics of the inhomogeneous electron-plasma system under consideration.

Our paper is organized as follows. In Section 2, we formulate the mathematical model and describe the numerical simulation technique. In Section 3, we discuss the nonlinear dynamic of the system. An introduction to the wavelet bicoherence analysis is given in Section 4, which summarizes the basic notions of both wavelet analysis and wavelet bicoherence analysis. In Section 5, the wavelet bicoherence analysis is employed to describe the dynamics of spatial structures by resolving them into coherent structures.

## 2. DESCRIPTION OF THE MODEL

We consider a diode gap between two plane electrodes held at the same potential and separated by distance $L$. A region of length $\Delta x_i$ with a boundary at $x_i$ is filled with neutralizing immobile ions of density $n_i$. An electron beam with space charge density $\rho_0 = n_0 e$ is injected at constant velocity $v_0$ into the diode gap. The control parameters of the system are as follows: the Pierce parameter

$$\alpha = \omega_p L / v_0, \tag{2}$$

which is the unperturbed transit angle in terms of the plasma frequency and depends on the beam current $I$ as $\alpha \sim \sqrt{I}$; the relative ion density

$$n = n_i / n_0 \tag{3}$$

and the geometric parameters $x_i$ and $\Delta x_i$ of the background ion layer. The last was assumed to be fixed and was set equal to $0.2L$. Note that, in the case $n = 1$, $x_i = 0$, and $\Delta x_i = L$ (a "classical" Pierce diode), the instability in the beam takes place at $\alpha > \pi$, in which case a virtual cathode forms in the system [5, 26].

We represent the beam as a system of charged plane sheets and write out the following set of dynamic equations for the space charge in the diode gap.

We introduce dimensionless variables, which are marked by a prime and are related to the dimensional variables by

$$\rho = \rho_0 \rho', \quad v = v_0 v', \quad \phi = v_0^2 \phi'/\eta,$$
$$x = Lx', \quad t = Lt'/v_0.$$

For each $j$th sheet, we solve the nonrelativistic equations of motion

$$\frac{dx_j}{dt} = v_j, \quad \frac{dv_j}{dt} = \left.\frac{\partial \phi}{\partial t}\right|_{x = x_j}, \tag{4}$$

where $x_j$ and $v_j$ are the coordinate and velocity of the $j$th sheet, $\phi(x, t)$ is the space-charge potential, and the primes for the dimensionless variables are omitted.

In order to determine the space-charge field, we solve the following Poisson's equation in the electrostatic approximation:

$$\frac{\partial^2 \phi}{\partial x^2} = \alpha^2 [\rho(x) - \rho_i(x)], \tag{5}$$

the boundary conditions being $\phi(0) = \phi(1) = 0$. Here, $\rho(x)$ is the space charge density profile, determined by the particle-in-cell method [27]. The function $\rho_i(x)$ describes the profile of the neutralizing charge of an inhomogeneous immobile ion background along the

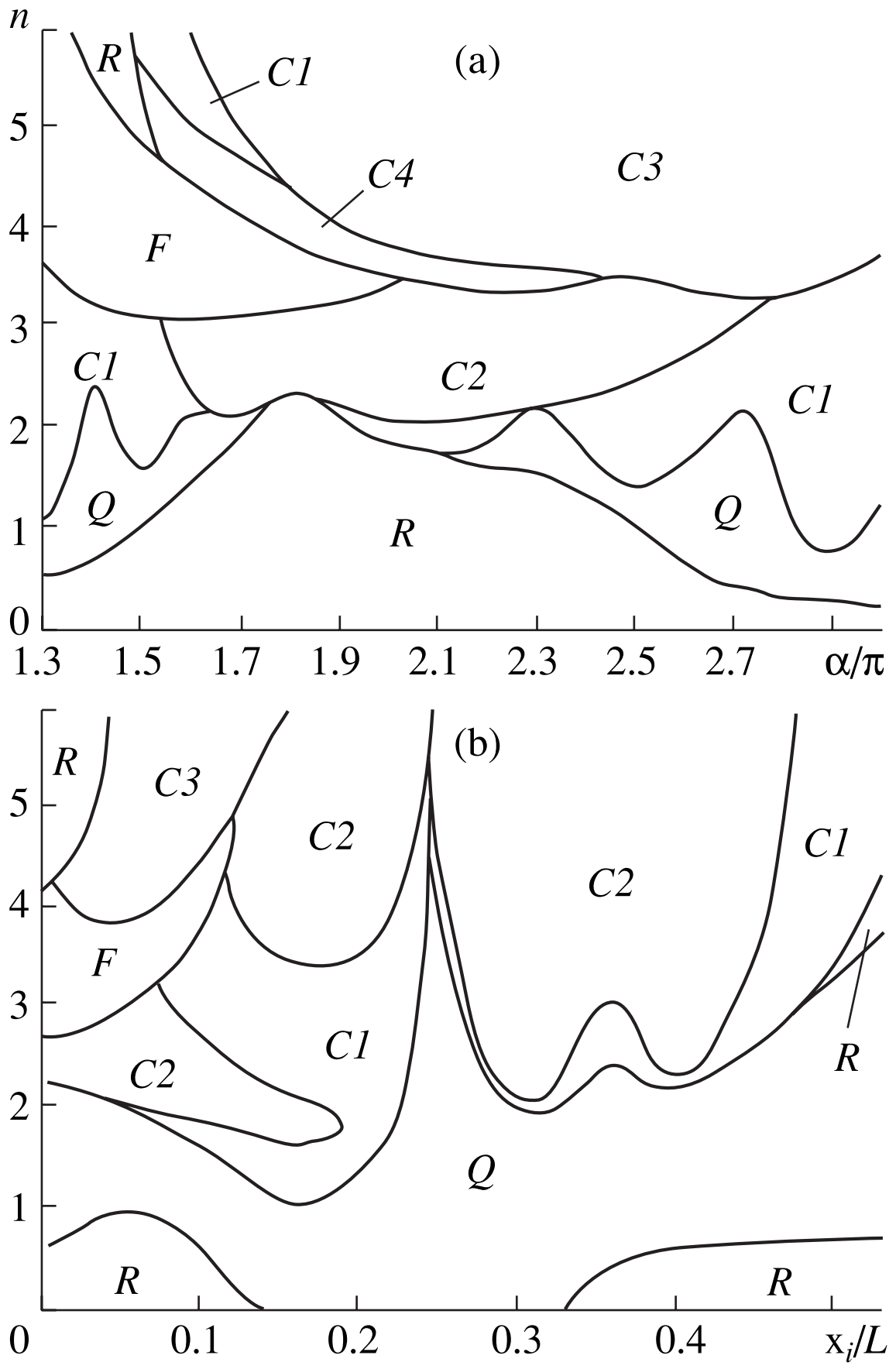

**Fig. 1.** Domains of the characteristic oscillation regimes in a diode gap with an overcritical current and an inhomogeneous ion background in the planes of control parameters: (a) the plane of the beam current α and the background ion density $n$ at $x_i/L = 0.05$ and (b) the left boundary $x_i$ of the ion layer and its density $n$ at $\alpha = 1.75\pi$.

system. In the case under analysis, the profile has the form

$$\rho_i(x) = \begin{cases} ne & \text{for} \quad x \ge x_i \quad \text{and} \quad x \le x_i + \Delta x_i, \\ 0 & \text{for} \quad x < x_i \quad \text{and} \quad x > x_i + \Delta x_i. \end{cases} \qquad (6)$$

The main parameters of the numerical scheme (see, e.g., [28]), such as the number $N_C$ of mesh points in the spatial grid and the number $N_0$ of particles in each cell of the grid in an unperturbed state, were chosen to be $N_C = 800$ and $N_0 = 24$ (in which case the total number of particles in the entire computation region in the unperturbed state was $N = 19200$). The above parameter values of the numerical scheme were chosen to achieve the desired accuracy and reliability of a numerical analysis of the complicated nonlinear processes (including deterministic chaos) in the electron-plasma system under investigation (see, e.g., [27, 29]).

The equations of motion were solved by a leap-frog scheme [27, 30], and Poisson's equation was integrated by the error-correction method [31].

## 3. NONLINEAR DYNAMICS OF THE SYSTEM

Here, we consider the nonlinear dynamics of the system when the control parameters—the Pierce parameter α and the parameters $n$ and $x_i$ of the background ion layer—are varied. The characteristic dynamic regimes in the system are presented in Fig. 1, which shows the domains corresponding to different regimes in the planes (α, $n$) and ($x_i$, $n$) of the control parameters.

The dynamic regimes were determined from the oscillations of the potential $\phi(x, t)$ in the cross section $x = x_\phi = 0.25L$. In order to reveal the oscillation regimes, the times evolutions of the potential $\phi(x_\phi, t)$ were calculated for different values of the control parameters. The calculated time realizations were used to obtain the oscillation power spectra and reconstruct the phase portrait of the oscillations. In turn, the phase portraits were reconstructed by Takens' method [32], which reduces to the construction of the phase vectors of the form

$$\mathbf{R} = \{\phi(t), \phi(t-T), \ldots \phi[t-(d-1)T]\}, \qquad (7)$$

where $\phi(t)$ are the oscillations of the potential in the cross section $x = x_\phi$, $T$ is the time delay, and $d$ is the dimensionality of embedding space ($d = 2$ corresponds to the projection of an attractor onto a plane). Figure 2 illustrates the results obtained for different possible dynamic regimes in the system.

An analysis of both the patterns of dynamic regimes (Fig. 1) and the parameters of oscillations of the potential (Fig. 2) shows that, when the beam current α, the background ion density $n$, and the position $x_i$ of the ion layer are varied, the oscillatory dynamics of the system undergoes complicated changes.

For low ion densities ($n < 1.0$–$2.0$), the oscillations of the virtual cathode are regular (regime $R$ in Fig. 1). (This regime is illustrated by Fig. 2a, which was obtained for the control parameters $\alpha = 1.75\pi$, $n = 0.5$, and $x_i = 0.05L$.) An increase in the ion density in the layer leads to the onset of quasi-periodic motion (regime $Q$), illustrated by Fig. 2b ($\alpha = 1.34\pi$, $n = 1.0$, and $x_i = 0.05L$). As the ion layer is displaced toward the right boundary of the system, the domain of quasi-periodic dynamics in the plane of the control parameters broadens.

A further increase in $n$ gives rise to different types of stochastic dynamics of electron waves in a diode gap with an inhomogeneous ion background. Domain $C1$ corresponds to the regime of weakly stochastic oscillations (see Fig. 2c; $\alpha = 2.86\pi$, $n = 2.0$, and $x_i = 0.05L$), which arise as a result of the disruption of quasi-periodic motion. In this case, the power spectrum is seen to peak at two incommensurable base frequencies, which are "inherited" from the quasi-periodic regime. The attractor is now a narrow strip in phase space. As the background ion density $n$ increases even further, regime *C1* evolves into stochastic regime *C2*, in which the oscillations in the system are more irregular and, as a consequence, the noisy pedestal in the power spectrum is higher. The attractor in phase space is a wide strip (see Fig. 2d; $\alpha = 2.0\pi$, $n = 2.5$, and $x_i = 0.05L$). At $x_i < 0.2$ and for a narrow range of values of the control parameters corresponding to regime *C4* in Fig. 1, developed stochastic oscillations are observed, whose characteristics are given in Fig. 2e ($\alpha = 2.0\pi$, $n = 3.5$, and $x_i = 0.05L$). In the power spectrum, there is a high noisy pedestal, decreasing gradually toward higher frequencies, and there are peaks at the base frequencies and at their satellite harmonics. The phase portrait of oscillations is uniform and is seen to be very noisy.

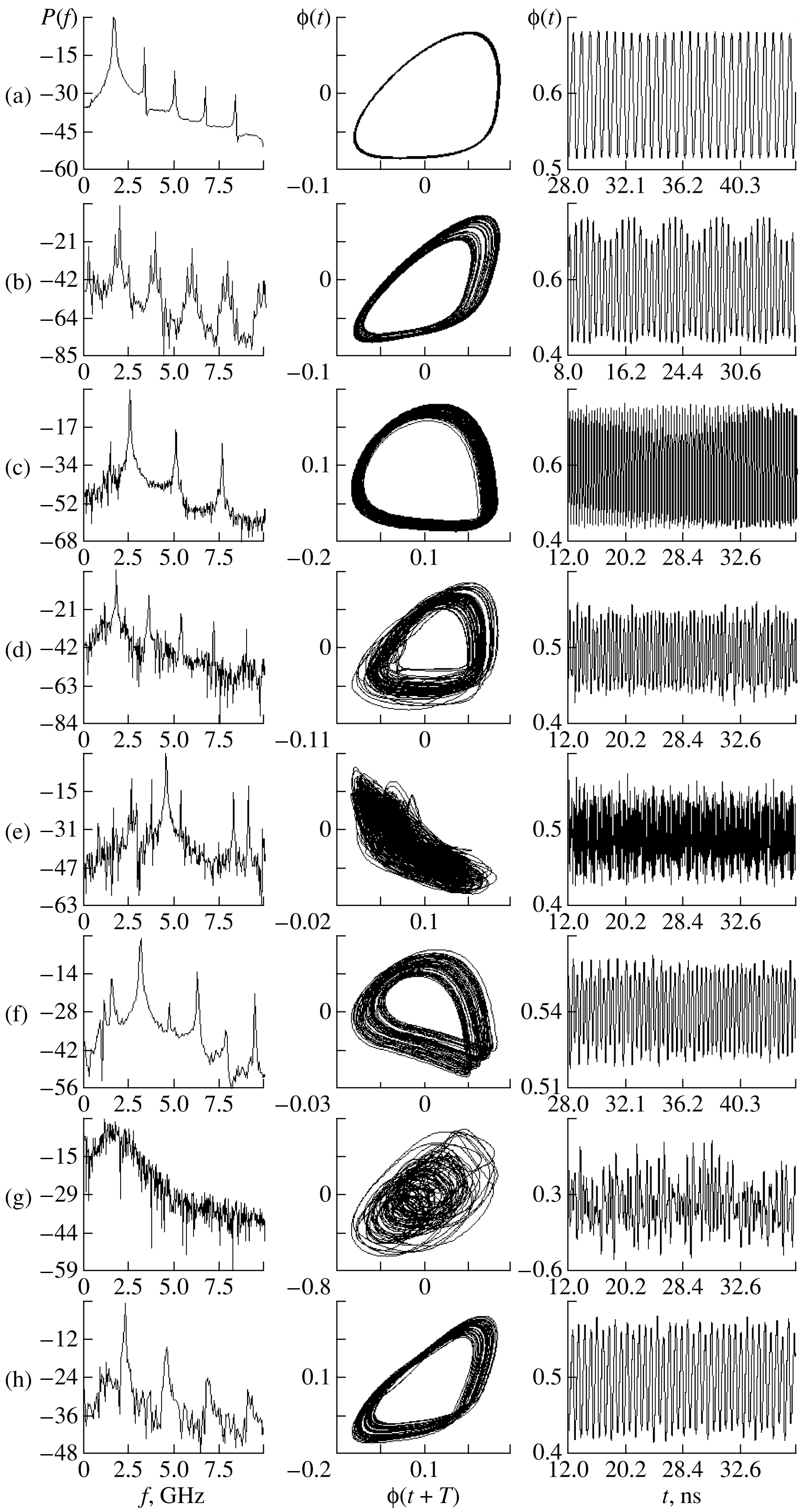

**Fig. 2.** Power spectra, phase portraits, and time realizations of oscillations of the potential $\phi(t)$ in the cross section $x/L = 0.25$ for different dynamic regimes possible in the system.

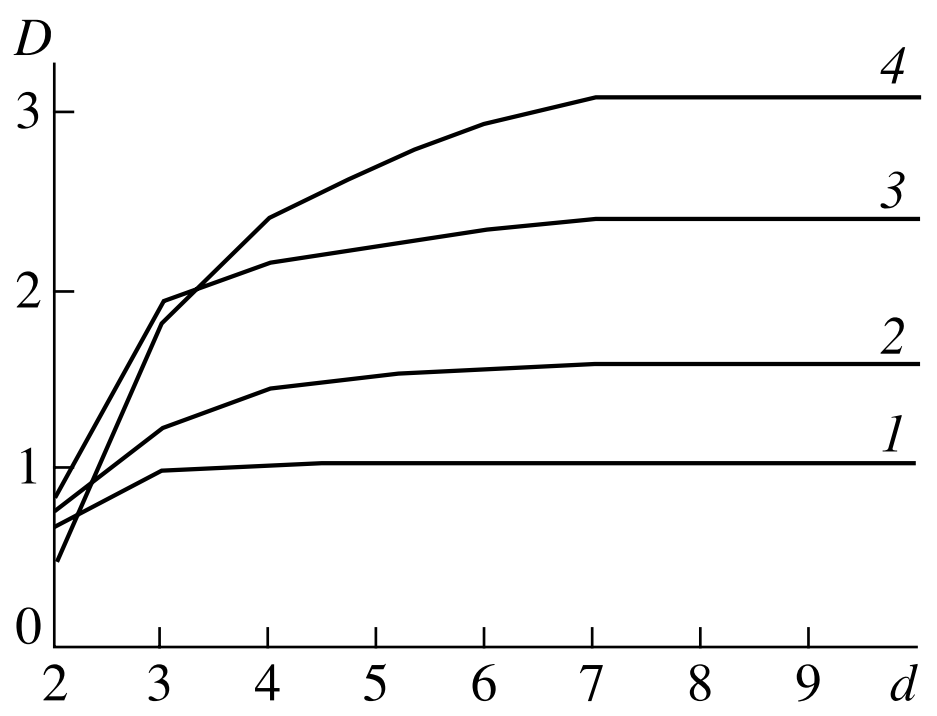

**Fig. 3.** Correlation dimensionality $D$ of an attractor as a function of the dimensionality $d$ of embedding space in different generation regimes: curves *1*, *2*, *3*, and *4* refer to regimes *C1*, *C2*, *C3*, and *C4*, respectively.

In order to quantitatively analyze the extent to which the stochastic regimes differ from each other, we determined the correlation dimensionality $D$ of stochastic attractors using the algorithm developed by Grassberger and Procaccia [33] (see also [34]). The correlation dimensionality $D$ of an attractor is a function of the observation scale length ε:

$$D(\varepsilon) = \lim_{\varepsilon \to 0} \frac{\ln C(\varepsilon)}{\ln \varepsilon}. \tag{8}$$

Here, the reduced correlation integral $C(\varepsilon)$ is defined as the number of pairs of points the distance between which in phase space is shorter than ε,

$$C(\varepsilon) = \frac{1}{MN} \sum_{j=1}^{M} \sum_{i=1,\, j \neq j}^{N} H(\varepsilon - |\mathbf{x}_i - \mathbf{x}_j|); \tag{9}$$

where $M$ is the number of reduction points, $N$ is the number of points in a time realization, $H$ is the Heaviside function, and **x** is the state vector in phase space (it was reconstructed by Takens' method). In simulations, the number of points in a time realization was $N$ = 200000 and the number of reduction points was $M$ = 10000. In this case, the correlation dimensionality is also a function of the dimensionality $d$ of embedding space in which the phase vector **x** is reconstructed [see formula (7)].

Figure 3 shows how the correlation dimensionality of an attractor depends on the dimensionality $d$ of embedding space in different characteristic oscillation regimes in the system under investigation. Note that, in all regimes, the dimensionality of the attractors saturates as the dimensionality $d$ of embedding space increases. This indicates that stochastic processes in a diode gap with an overcritical current and an inhomogeneous ion background are of a dynamic nature, because, in the case of noisy oscillations (including those introduced by numerical errors), the dimensionality $D$ has a tendency to increase in proportion to the dimensionality $d$ of embedding space [35, 36].

In the regime of weakly stochastic oscillations (regime *C1*), the dimensionality of the attractor of stochastic oscillations, which is a narrow strip in phase space, is equal to $D_{C1}$ =1.04 ± 0.01. We see that the dimensionality is fractional, which is typical of dynamic chaos. The small value of the dimensionality in regime *C1* indicates that the attractor in phase space is a very narrow strip, whose dimensionality differs insignificantly from the dimensionality $D_R = 1.0$ of the attractor of regular oscillations (limiting cycle), $D_{C1} - D_R = 0.04$. However, as the oscillations become more irregular (which corresponds to a transition to stochastic regime *C2*), the dimensionality of the stochastic attractor increases up to $D_{C2} = 1.62 \pm 0.01$. This indicates that the oscillations are of a very complex nature: the attractor is now a wide strip in phase space, $D_{C2} - D_R = 0.62$. In the regime of developed stochastic oscillations (regime *C4*), the dimensionality of the stochastic attractor is equal to $D_{C4} = 3.12 \pm 0.02$, which is much larger than that in the preceding regimes.

In the plane of control parameters, we can also distinguish a "beak," which is denoted by $F$ and in which the oscillation frequency in an electron beam increases sharply. The corresponding oscillation regime is illustrated by Fig. 2f ($\alpha = 1.75\pi$, $n = 3.5$, and $x_i = 0.05L$), which was obtained for the same beam current as in Fig. 2a, but for a substantially denser ion layer. A comparison of the power spectra in these two figures shows that the base generation frequency increases by a factor of ~2.

The dependence of the base frequency in the oscillation spectrum on both the beam current and the ion density in the layer, calculated at $x_i/L = 0.05$, is presented in Fig. 4.

From Fig. 4a, which shows the dependence of the base frequency on the Pierce parameter (the beam current) at different densities $n$, we can see that, for low ion densities in the layer ($n$ = 0.5, curve *1*), the frequency is a linear function of α. This result is quite natural: in a diode gap with a homogeneous ion background, the oscillation frequency $f$ of the virtual cathode is determined by the Langmuir frequency $\omega_p$ of the electron beam, $f \approx \omega_p/\pi$ (see, e.g., [37–39]). From definition (2) of the Pierce parameter, it is clear that, for low ion densities in the layer, the oscillation frequency is $f \sim \alpha$, which agrees with the results of our numerical experiments. As the ion density increases ($n$ = 2.5, curve *2*), the dependence $f(\alpha)$ turns from being linear; moreover, with increasing Pierce parameter, the frequency becomes progressively lower than that in the case of low ion densities in the layer (curve *2* deviates gradually downward from curve *1*). As the ion density increases further, the situation changes radically ($n$ = 3.5, curve *3*). For $\alpha < 2.1\pi$, the oscillation frequency in the beam at high ion densities in the layer is substantially higher than at low values of $n$. Note that, at $\alpha \sim 2.1\pi$, the oscillation frequency first decreases and then increases abruptly to $f \approx 6$ GHz. For even higher beam currents (larger α values) and for high $n$ values, the base oscillation frequency in the spectrum decreases sharply.

This is also illustrated in Fig. 4b, which displays the dependence $f(n)$ for different beam currents (different α values) from domain $F$ in the pattern of regimes shown in Fig. 1a. From Fig. 4b, we can see that, as the density $n$ reaches a certain critical value $n_{cr1}$, the base frequency in the oscillation spectrum increases abruptly by a factor of ~2. The higher the beam current, the higher is the critical density $n_{cr1}$. This is distinctly seen from a comparison of the curves calculated for different values of the Pierce parameter α.

As the ion density $n$ in the layer increases further, the frequency $f(n)$ first somewhat decreases, reaching its minimum values, and then increases in a jumplike manner. In this case, the frequency is higher than the oscillation frequency for $n \gtrsim n_{cr1}$. For even higher ion densities ($n_{cr2} > n_{cr1}$), the oscillation frequency in an electron beam decreases sharply. In contrast to $n_{cr1}$, the critical density $n_{cr2}$ decreases with increasing Pierce parameter; in other words, as the beam current increases, the region in which the oscillation frequency in an electron beam with a virtual cathode increases becomes narrower and, simultaneously, the oscillation frequency at ion densities in the range $n \in (n_{cr1}, n_{cr2})$ increases.

Note also that, in domain *C4* of the control parameters, the most intense peak in the oscillation spectrum occurs at the base oscillation frequency in regime *F*.

For very high ion densities in a layer located close to the injection plane for an electron beam, the oscillations excited in the system are highly irregular (regime *C3*): there is no base frequency and the spectral power decreases sharply toward higher frequencies (Fig. 2g; $\alpha = 2.4\pi$, $n = 5.0$, and $x_i = 0.05L$). In this case, the phase portrait of oscillations is uniform, and, as the dimensionality $d$ of embedding space increases, the dimensionality of the attractor saturates at $D_{C3} = 2.43 \pm 0.02$ (Fig. 3).

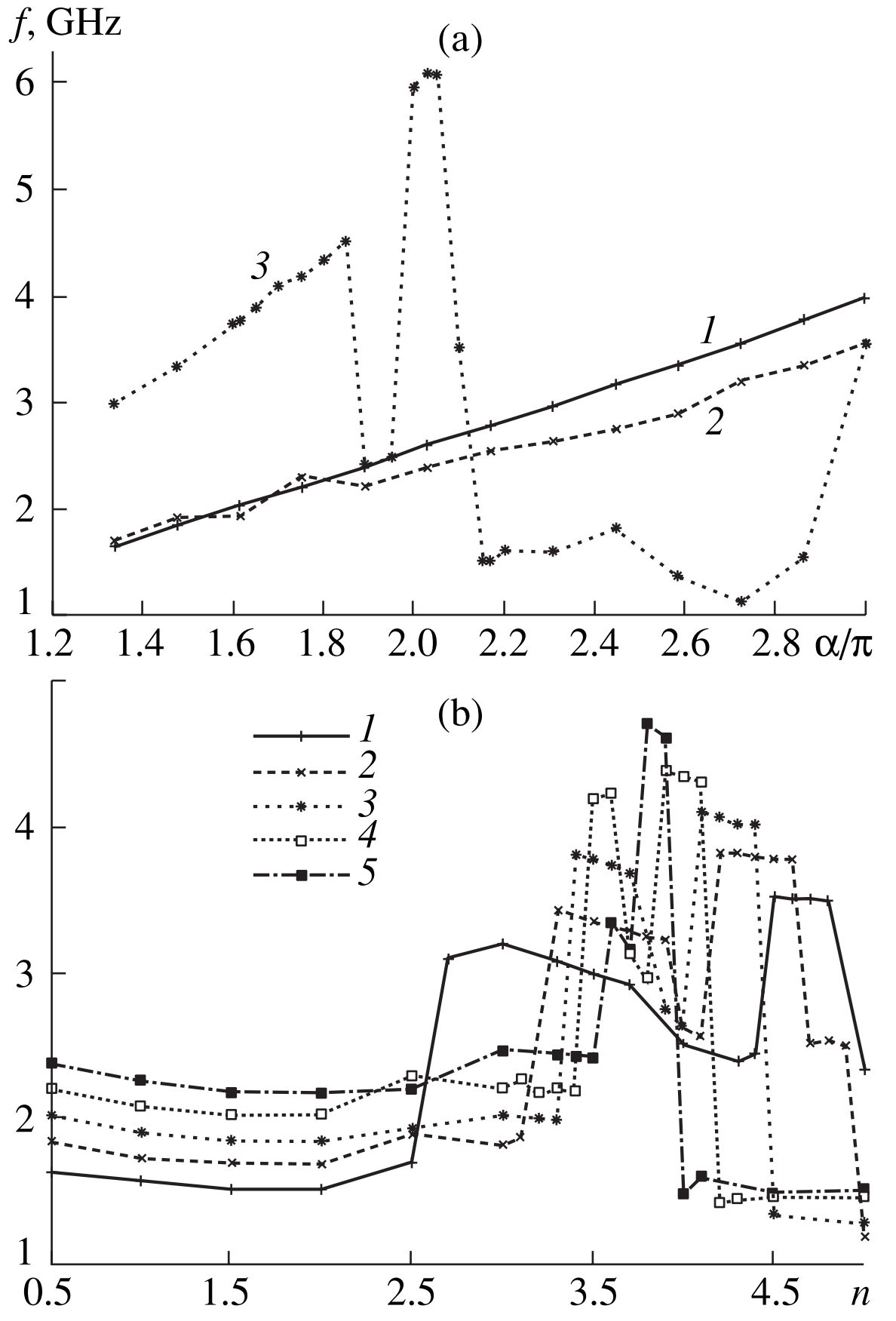

**Fig. 4.** (a) Oscillation frequency vs. Pierce parameter α for different values of $n$: $n$ = (*1*) 0.5, (*2*) 2.5, and (*3*) 3.5; (b) oscillation frequency vs. ion density $n$ in the layer for different values of the Pierce parameter: α = (*1*) 1.34π, (*2*) 1.48π, (*3*) 1.61π, (*4*) 1.75π, and (*5*) 1.89π.

However, for high ion densities ($n > 3$) in a layer located far from the injection plane for an electron beam ($x_i/L > 0.2$), the system exhibits far simpler dynamics near the injection plane (Fig. 2h; $\alpha = 3.0\pi$, $n = 4.5$, and $x_i = 0.25L$). The phase portrait and the shape of the power spectrum are seen to be similar to those in regime *C2* (Fig. 2d); this circumstance allows the observed dynamic regime to be classified as *C2* regime.

Let us consider the causes of such complicated transitions between different oscillation regimes as the control parameters of the system—the beam current and the inhomogeneity of the background ion layer—are varied. Following papers [10, 11, 13, 14], which have been cited in the Introduction, we can suggest that the above behavior of the system is governed by the distinctive features of the dynamics and interaction of coherent structures that form in the system. In order to examine the processes of the formation and interaction of coherent structures in an electron beam, we turn to the wavelet bicoherence analysis.

## 4. WAVELET ANALYSIS AND WAVELET BICOHERENCE

A continuous wavelet transformation is defined as the convolution

$$W(s, t_0) = \int_{-\infty}^{+\infty} g(t)\psi^*_{s, t_0}(t)dt \qquad (10)$$

of the signal $g(t)$ to be analyzed with the two-parameter wavelet function $\psi_{s, t_0}(t)$, which is, in turn, defined in terms of the mother wavelet $\psi_0(t)$ as

$$\psi_{s, t_0}(t) = \frac{1}{\sqrt{s}}\psi_0\left(\frac{t-t_0}{s}\right), \qquad (11)$$

where the asterisk stands for complex conjugation. The parameter $s$, which is called the scale of the wavelet transformation, describes the wavelet duration, and the displacement parameter $t_0$ determines the position of the wavelet at the time axis $t$. In definition (11), the factor $1/\sqrt{s}$ is introduced in order for all of the wavelet functions $\psi_{s, t_0}$ to be normalized to unity, $\int_{-\infty}^{+\infty}(\psi_{s, t_0})^2 dx = 1$. The role of the wavelet function can be played by any function satisfying the following three main conditions.

(i) *The localization condition*, which indicates that the mother wavelet $\psi_0$ should be localized both in time and along the frequency axis. To satisfy this condition, it is enough to choose a sufficiently regular function $\psi_0$ defined over a finite interval.

(ii) *The admissibility condition*, which indicates that the mother wavelet should be chosen so that its Fourier transform $\hat{\psi}_0(\omega)$ satisfies the condition

$$C_\psi = 2\pi\int_{-\infty}^{+\infty}\frac{|\hat{\psi}_0(\omega)|^2}{\omega}d\omega < \infty. \qquad (12)$$

For all practical purposes, condition (12) is equivalent to the *zero-mean condition*

$$\int_{-\infty}^{+\infty}\psi_0(t)dt = 0 \quad \text{or} \quad \hat{\psi}_0(0) = 0. \qquad (13)$$

Note that continuous wavelet transformation (10) is an expansion in the nonorthogonal basis $\psi_{s, t_0}$. However, under admissibility condition (12), there exists the reverse wavelet transformation

$$g(t) = \frac{1}{C_\psi}\int_0^{+\infty}\frac{ds}{s^2\sqrt{s}}\int_{-\infty}^{+\infty}\psi_0\left(\frac{t-t_0}{s}\right)W(s, t_0)dt_0. \qquad (14)$$

The bicoherent wavelet transformation provides a tool for calculating the wavelet bispectrum and is a generalization of the wavelet transformation. The normalized bispectrum (bicoherence) characterizes the phase relations (phase coupling) between different frequency components of the signal spectrum. It is possible to speak of the phase coupling if the signal to be analyzed contains two components such that the sum (or difference) of their frequencies $\omega_1$ and $\omega_2$ and the sum of their phases $\phi_1$ and $\phi_2$ both remain constant during a certain time interval. Bicoherence is a quantitative measure of this phase coupling. The value of bicoherence is a function of the frequencies $\omega_1$ and $\omega_2$ and should be close to unity if the signal to be analyzed contains three components whose frequencies $\omega_1$, $\omega_2$, and $\omega$ and phases satisfy the relationships

$$\omega_1 + \omega_2 = \omega, \quad \phi_1 + \phi_2 = \phi + \text{const}. \qquad (15)$$

If relationships (15) fail to hold, bicoherence between the components with frequencies $\omega_1$, $\omega_2$, and $\omega$ is equal to zero.

The reason for this is that, in determining the Fourier or the wavelet spectrum at any time, the process under analysis is regarded as a superposition of statistically uncorrelated processes occurring on different time scales and an estimate is made of how the power is distributed between these scales. In this case, the dynamics of the process is described in terms of only linear mechanisms, because the phase relations between different frequency components of the spectrum are ignored. In fact, we can say that the Fourier or the wavelet spectrum contains enough information to provide a complete statistical description of a random Gaussian process with a known mean value. However, when the signal to be analyzed is generated by a *nonlinear* dynamic system and, accordingly, is a complicated superposition of different frequency harmonics at any time, the information contained in the Fourier or the wavelet spectrum is insufficient to provide a complete description of the process. It can thus be suggested that the phase coupling appears between some of the frequency components of the signal. Information concerning the nonlinearities makes it possible to obtain the higher order spectra, in particular, the third-order spectrum, or bispectrum. Extending the bispectral analysis to include the wavelet transformation makes it possible to analyze essentially nonlinear phenomena (such as the temporal dynamics of the phase coupling between different signal components) and to resolve coherent structures in spatiotemporal data.

By analogy with the Fourier bispectrum [25, 40], the wavelet mutual bispectrum can be defined as [16]

$$B_{hg}(s_1, s_2) = \int_T W_h^*(s, \tau) W_g(s_1, \tau) W_g(s_2, \tau) d\tau, \quad (16)$$

where

$$\frac{1}{s} = \frac{1}{s_1} + \frac{1}{s_2} \quad \text{or} \quad \omega_s = \omega_{s1} + \omega_{s2} \quad (17)$$

is the rule for summing the frequencies and the frequency $\omega_s = 2\pi/s$ corresponds to the time scale $s$ of the wavelet transformation.

The wavelet mutual bispectrum is a measure of the phase coupling (over the time interval $T$) between the components of the wavelet spectrum $W$ of the signal $g(t)$ on scales $s_1$ and $s_2$ and the component of the same spectrum of the signal $h(t)$ on scale $s$. Accordingly, the wavelet mutual bispectrum can be interpreted as the phase coupling between waves (wavelets) whose frequencies satisfy the second of conditions (17).

For ease of analysis, it is convenient to use not only mutual bispectrum (16) but also the wavelet mutual bicoherence, which is defined to be the normalized wavelet mutual bispectrum:

$$b_{hg}(s_1, s_2) = |B_{hg}(s_1, s_2)| \times \left( \int_T |W_g(s_1, \tau) W_g(s_2, \tau)|^2 d\tau \int_T |W_h(s, \tau)|^2 d\tau \right)^{-0.5}. \quad (18)$$

According to this definition, the values of the mutual bicoherence lie in the unit segment [0, 1].

An important characteristic of the bicoherent wavelet transformation is the summed bicoherence, which is defined as

$$b_\Sigma(s) = \sqrt{\frac{1}{N_s} \sum [b(s_1, s_2)]^2}, \quad (19)$$

where the summation is over all scales $s_1$ and $s_2$ satisfying the first of conditions (17) and $N_s$ is the number of terms in the sum. Another important characteristic is the total bicoherence, defined as

$$b = \sqrt{\frac{1}{N_S} \sum \sum [b(s_1, s_2)]^2}, \quad (20)$$

where the sum is now extended over all scales $s_1$ and $s_2$ to be analyzed and $N_S$ is the total number of quantities to be summed. The coefficients $1/N_s$ and $1/N_S$ are required in order that the quantities $b_\Sigma$ and $b$ to have values lying in the unit segment [0, 1].

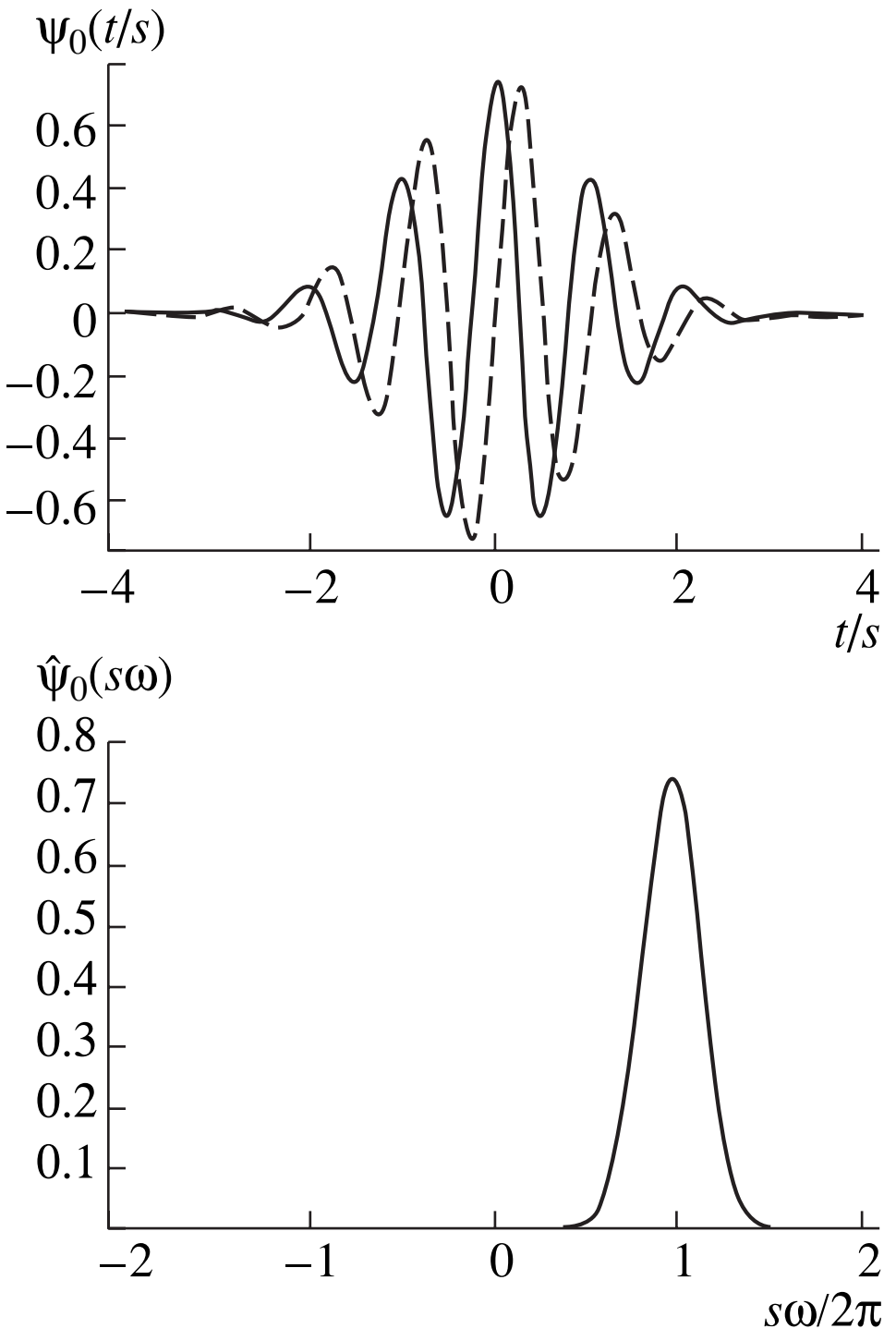

**Fig. 5.** Shapes of Morlet's wavelet $\psi_0$ with $\omega_0 = 6$ and its Fourier transform $\hat{\psi}_0$. The solid and dashed curves are for the real and imaginary parts of the wavelet, respectively.

The summed bispectrum can be conveniently reconstructed from the numerically obtained wavelet mutual bispectrum:

$$B_\Sigma(s) = \sqrt{\frac{1}{N_s} \sum [B(s_1, s_2)]^2}, \quad (21)$$

where the notation is the same as in formula (19).

Recall that the mother wavelet may be chosen ambiguously. The question then arises as to the wavelet function $\psi_0(\eta)$ that should be used to calculate the wavelet bicoherence and wavelet mutual bispectrum. In our opinion, the most convenient for further interpretation of the results is Morlet's complex wavelet [41]

$$\psi_0(\eta) = \pi^{-1/4} \exp(j\omega_0\eta) \exp(-\eta^2/2), \quad (22)$$

where $\omega_0$ is the wavelet parameter. The usual practice is to consider Morlet's wavelet with $\omega_0 = 6$. The basis of the Morlet's wavelet is well localized in both coordinate and frequency spaces; moreover, the larger the parameter $\omega_0$, the higher is the resolution in frequency space and the worse is the localization in time. In fact, Morlet's wavelet is a sinusoidal function modulated by the Gaussian function (see Fig. 5, which shows the shapes of Morlet's wavelet and its Fourier transform). It is easy to see that there is a direct analogy between the scales $s$ (or the corresponding frequencies $\omega_s$) of the wavelet or the bicoherent wavelet transformation with the basic Morlet's wavelet and the periods $\Lambda$ (or the frequencies $\omega = 2\pi/\Lambda$) of the Fourier transform. This analogy makes it possible to examine the results of the wavelet bicoherence analysis in terms of the signal frequencies and phases, as is commonly done in Fourier analysis.

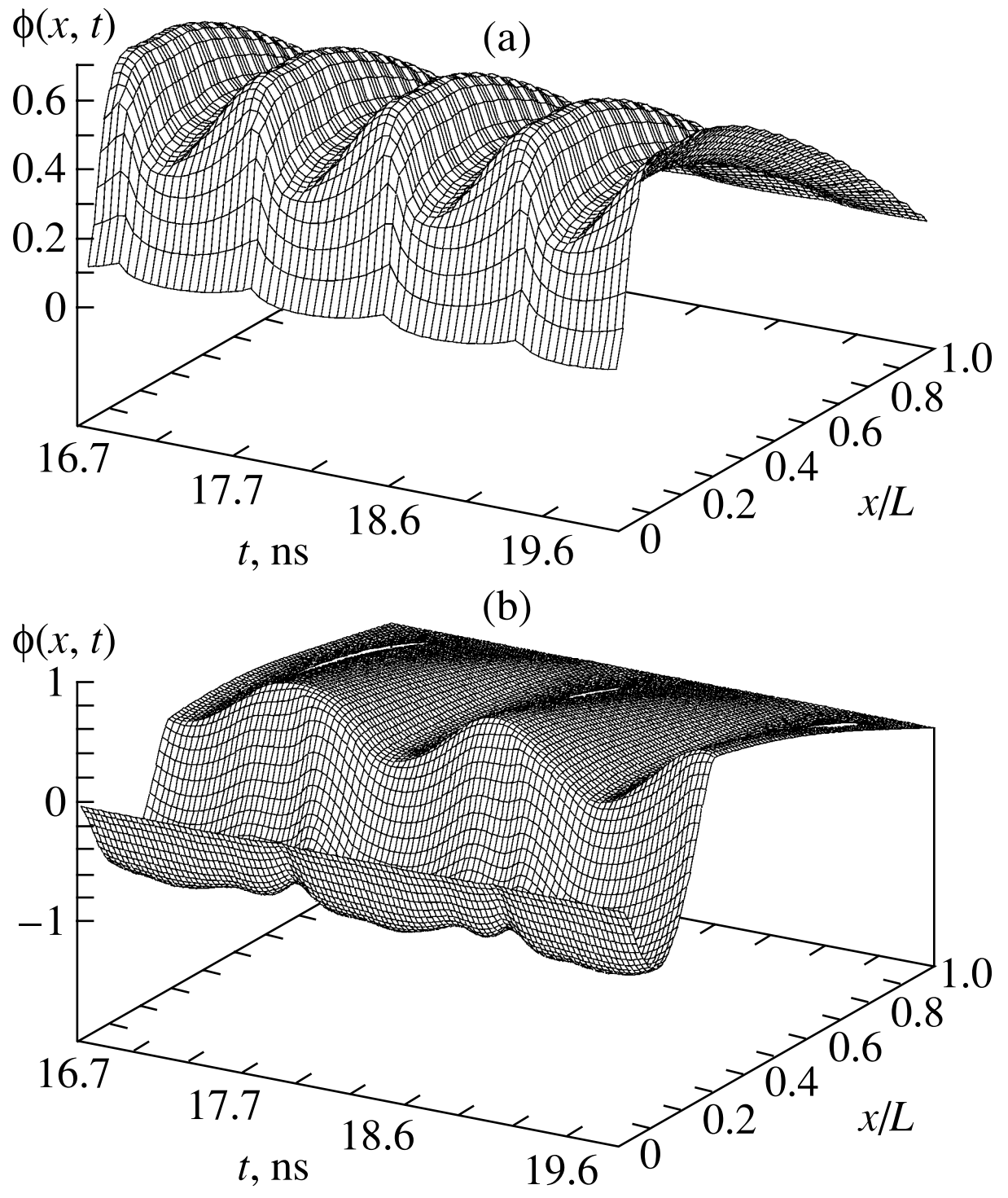

**Fig. 6.** Spatiotemporal dynamics of the space charge potential in the regimes of (a) regular oscillations (regime *R* in Fig. 1) and (b) stochastic oscillations (regime *C3*).

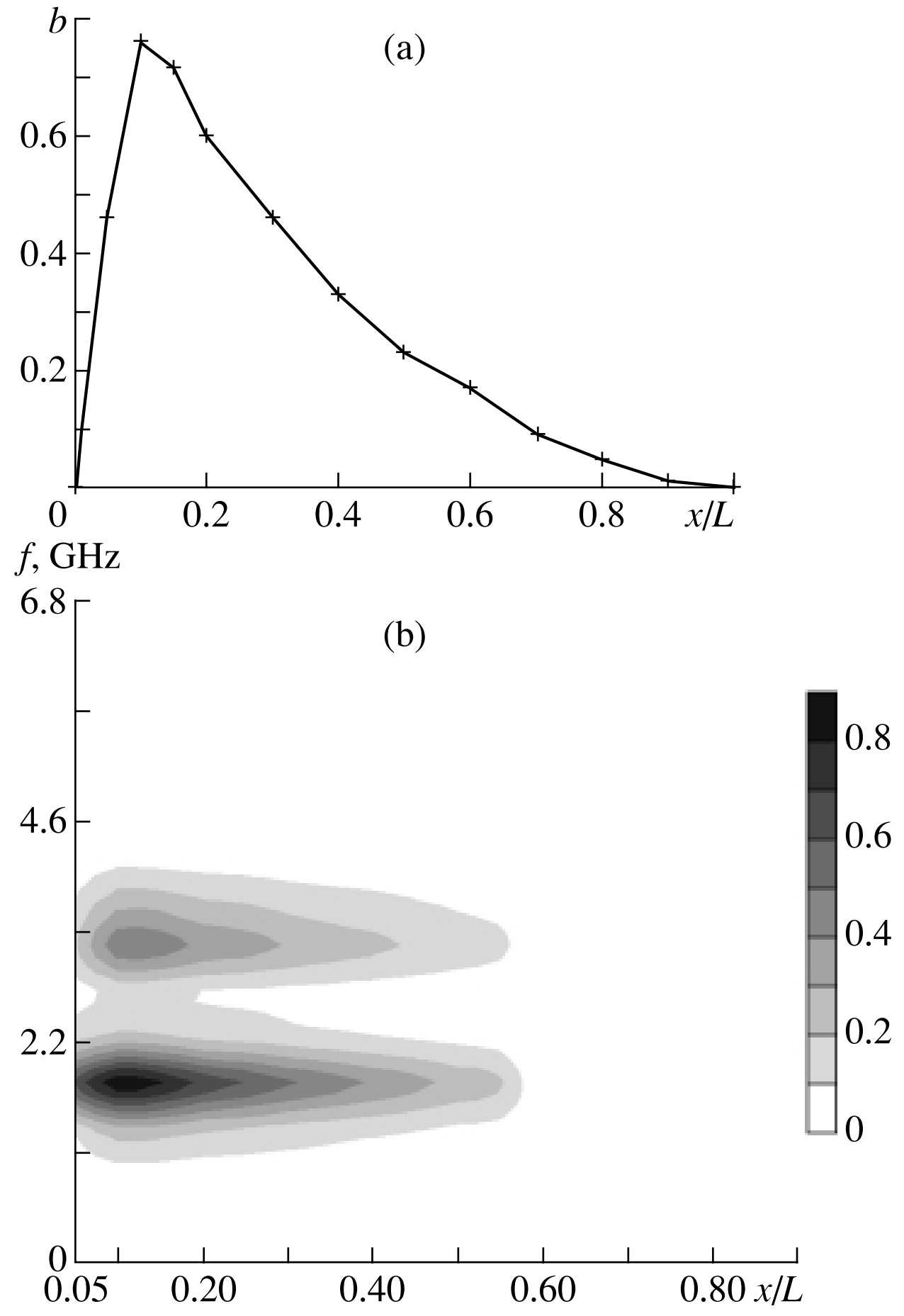

**Fig. 7.** (a) Coordinate dependence of the total bicoherence $b(x)$ and (b) the relief of the summed bispectrum $B_{\Sigma}(x/L, f)$ in the regime of regular oscillations (regime *R*).

## 5. COHERENT STRUCTURES

In order to analyze the processes occurring in an electron beam with an overcritical current by the methods of wavelet bicoherence, we investigate the spatiotemporal data on different characteristic regimes of oscillations of the potential $\varphi(x, t)$ of the space charge field in a diode gap.

We start by considering the simplest case—the regime of regular oscillations, which occurs at low beam currents $\alpha$ against a slightly inhomogeneous ion background (see domain *R* in Fig. 1). Figure 6a shows the corresponding spatiotemporal distribution of the potential $\phi(x, t)$ for the parameter values $\alpha = 1.54\pi$, $n = 0.5$, and $x_i = 0.05$. We can see that, in all cross sections of the diode gap, there are oscillations at the base frequency $f_0 = 0.98$ GHz.

Let us determine total bicoherence (20) of the data on oscillations of the potential $\phi(x, t)$ and of the harmonic signal whose frequency corresponds to the base frequency $f_0$ in the power spectrum of oscillations of the potential. In other words, we choose the signal $a\sin 2\pi f_0 t$ as the signal $h(t)$ in relationship (20), and we take the oscillations of $\phi(x_{\text{fix}}, t)$ at a certain fixed spatial point $x = x_{\text{fix}}$ as the signal $g(t)$. In calculations, as the mother wavelet function, we used Morlet's wavelet, which provides an easy interpretation of the results obtained (see the discussion above). Figure 7a shows the calculated coordinate dependence of the total bicoherence. The function $b(x)$ is seen to have one maximum, which occurs near the injection plane. This indicates that the highest possible bicoherence (i.e., the strongest phase coupling between the oscillations of the potential and the dynamics of the process on the main time scale) corresponds to the oscillations of the potential in the region $x/L \approx 0.1$. We thus can conclude that the main spatiotemporal structure, which determines the basic time and spatial scales of the dynamics of the system under investigation, is localized in the region $x/L \sim 0.1$.

Now, we calculate the summed wavelet bispectrum $B_{\Sigma}(f)$ [see relationship (21)] of the spatiotemporal data on oscillations of the potential in a diode gap. The oscillations of the potential $\phi(x, t)$ in the cross section $x/L$ = 0.1, in which the total bicoherence of the process is the highest, can be chosen as the signal $h(t)$, and the oscillations of $\phi(x, t)$ at different cross sections of the diode gap can be chosen as the signal $g(t)$. With these choices, it is possible to reveal the spatial regions of the strong phase coupling between time oscillations, i.e., to resolve the coherent structures that form in the system and to determine their spatial localization.

The results of calculating the summed wavelet bispectrum are illustrated in Fig. 7b, which shows the relief of the quantity $B_\Sigma$ on the coordinate plane $(x, f)$, where $x$ is the coordinate of the cross section with respect to which the bicoherence is calculated and the symbol $f$ denotes the frequencies for which the summed bicoherence is determined. We can see that, in the distribution of the summed bicoherence $B_\Sigma(x, f)$, there is a pronounced localized region where the wavelet bicoherence increases sharply, namely, the region $x/L \sim 0.05$–$0.2$ and $f \sim 0.8$–$1.2$ GHz in the coordinate plane $(x, f)$. This is the region of a single spatiotemporal structure in the beam. The structure is located near the injection plane (is spatially bounded by $x/L < 0.2$) and its characteristic time scale is about $T = 1/f \sim 1$ ns. Note that the distribution of the summed bicoherence has another peak at the frequency $f \sim 3.8$ GHz, which is approximately three times lower in height than the peak at the frequency $f \sim 1.9$ GHz. This second peak is associated with the dynamics of the same coherent structure and is determined by the quadratic nonlinearity of the electron-plasma system, specifically, by the phase coupling between the dynamic processes occurring on the main time scale $T$ and on the time scale $T/2$ of the second harmonic.

As the ion background becomes more inhomogeneous (as the density $n$ of the ion layer becomes higher), the generation frequency in the system increases abruptly (domain $F$ in Fig. 1). Let us analyze the dynamics of the system in this regime by the methods of wavelet bicoherence. We consider the oscillations of the potential $\phi(x, t)$ for the same beam current ($\alpha = 1.54\pi$) and the same position of the ion layer ($x_i$ = 0.05) as in the previous case, but for a substantially higher ion density ($n$ = 3.75). Figure 8a shows the corresponding coordinate dependence $b(x)$ of the total bicoherence between the oscillations of the potential in different cross sections $x$ of the diode gap and the harmonic signal at the fundamental generation frequency $f_0$ = 3.1 GHz, and Fig. 8b shows the distribution of the summed bicoherence $B_\Sigma(x, f)$.

From Fig. 8a, we can see that the region of high wavelet bicoherence is adjacent to the injection plane $x = 0$. The distribution of the amplitudes of the coefficients of the summed wavelet bispectrum along the interaction space (Fig. 8b) shows that, as before, there is only one coherent structure, which characterizes the behavior of an electron beam in the system. However, in the interaction space, the structure now forms near the injection plane of the diode gap, in the interval $x/L \in (0.0, 0.1)$. The characteristic time scale of this dynamic structure (approximately equal to $T \approx 0.5$ ns) is about two times longer than that in the previous case of a low ion density in the layer. As in the previous case, the temporal dynamics of the second harmonic of the base frequency is very pronounced, indicating that the oscillations are highly nonlinear.

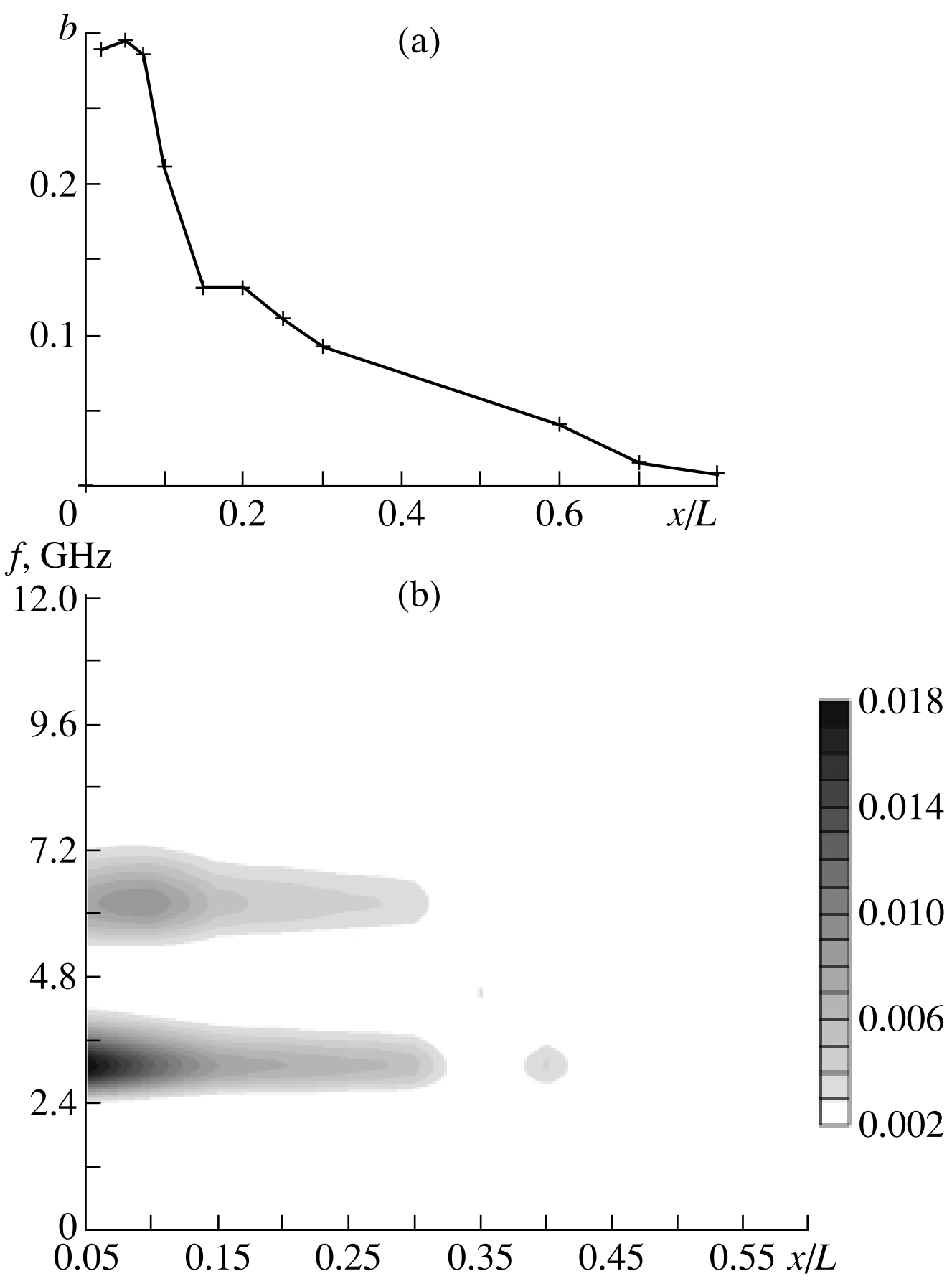

**Fig. 8.** (a) Coordinate dependence of the total bicoherence $b(x)$ and (b) the relief of the summed bispectrum $B_\Sigma(x/L, f)$ in regime $F$.

Hence, in the case under consideration, an increase in the degree of inhomogeneity of the system changes the conditions for the formation of the coherent structure governing the dynamics of an electron beam in the system. As a result of such a reconfiguration of the internal structure of the beam, the characteristic spatial scale $\Lambda_F$ of the dynamics of the single structure that forms in the system changes: it becomes about two times shorter than the spatial scale $\Lambda_R$ of the coherent structure in the regime of regular oscillations.

It is well known [42] that, for systems with a virtual cathode, the mean transit time $\tau_e$ of the electrons reflected from the virtual cathode is related to the oscil-lation frequency $f$ by the simple relationship $f \sim 1/\tau$. In the case at hand, the transit times in regimes $F$ and $R$ can be estimated as $\tau_F \sim \Lambda_F/v_0$ and $\tau_R \sim \Lambda_R/v_0$, respectively. These estimates show that the oscillation frequency $f_F$ in the $F$ regime in a diode with a highly inhomogeneous ion background is related to the generation frequency $f_R$ in a diode with a slightly inhomogeneous background by $f_F \approx 2f_R$, which is confirmed by the results of our numerical modeling.

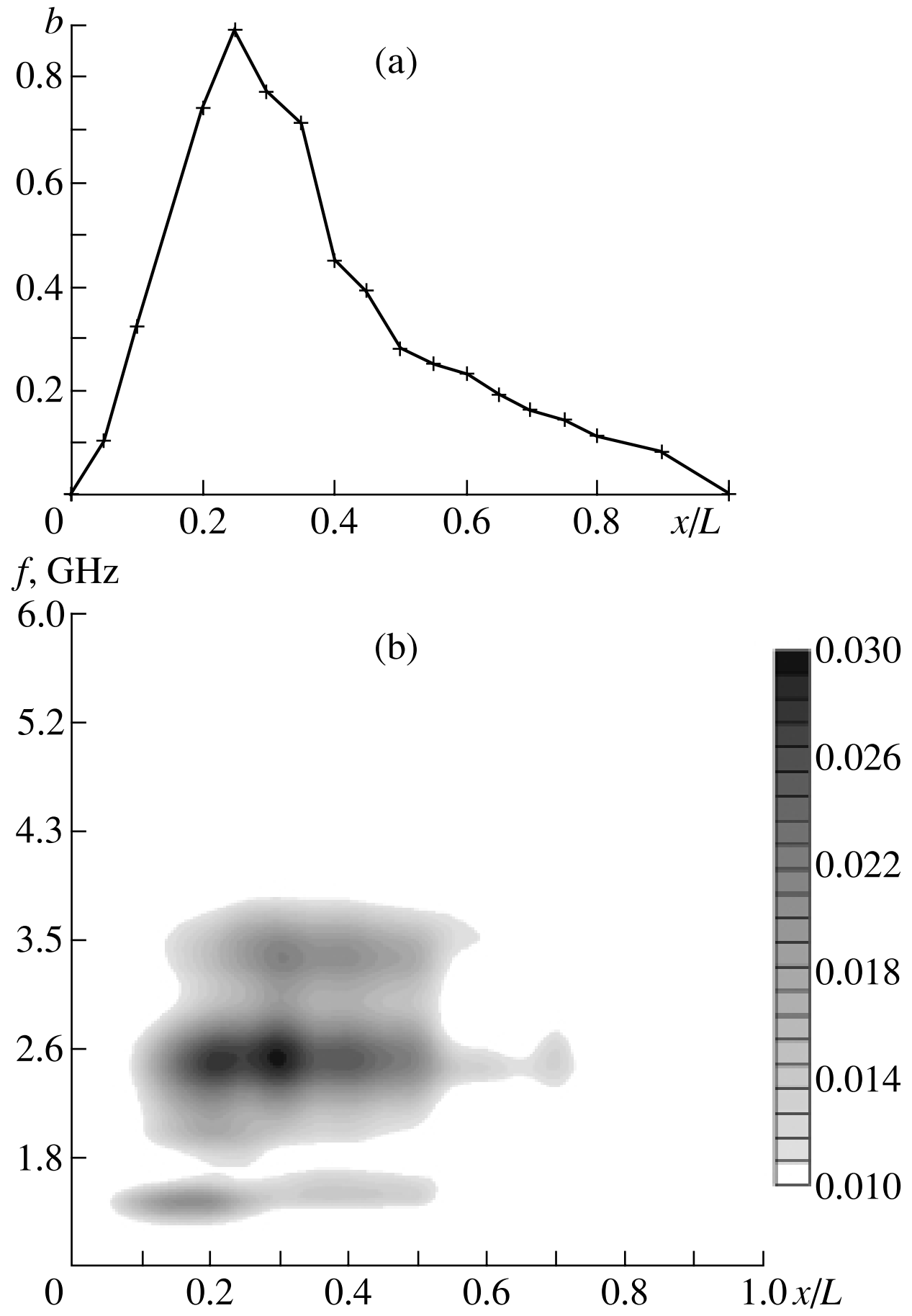

**Fig. 9.** (a) Coordinate dependence of the total bicoherence $b(x)$ and (b) the relief of the summed bispectrum $B_\Sigma(x/L, f)$ in regime *C3*.

Now, we will analyze the regimes of developed stochastic oscillations in the system.

We start by investigating the regime that arises at a high beam current $\alpha$ in a diode gap with a highly inhomogeneous background ion density $n$ and with an ion layer adjacent to the injection plane (domain *C3* in Fig. 1). We consider the spatiotemporal oscillations of the potential of the space charge field for the following values of the control parameters: $\alpha = 2.125\pi$, $n = 4.0$, and $x_i = 0.05$. The relevant distribution of the potential $\phi(x, t)$ is shown in Fig. 6b. Note that, in this regime, the spectral composition of the oscillations of the potential in different cross sections of the diode remains essentially the same.

Figure 9 illustrates the results of calculating the coordinate dependence of the total bicoherence $b(x)$ of oscillations of the potential and the distribution of the summed bicoherence $B_\Sigma(x, f)$. According to Fig. 9a, the highest total bicoherence corresponds to the oscillations of the potential in the region $x/L \approx 0.25$. This indicates that, in the regime at hand, the main spatiotemporal structure, which governs the characteristic dynamic features of the system in question (first of all, the fundamental spatial and time scales of oscillations in the beam), is localized in the region $x/L \sim 0.25$.

Our calculations of the summed wavelet bispectrum show that, in the distribution $B_\Sigma(x, f)$, there are two pronounced regions in the $(x, f)$ plane where the wavelet bicoherence increases sharply: the region $x/L \sim 0.2$–$0.4$ and $f \sim 2.5$ GHz and the region $x/L \sim 0.1$–$0.2$ and $f \sim 1.0$ GHz. In each of these regions (where the coefficients of the wavelet bispectrum are large), there is phase coupling between oscillations occurring in different cross section of the diode gap on the corresponding time scales. Moreover, the value of bicoherence in the first region is considerably larger than that in the second region. Each of these regions in the $(x, f)$ plane can be characterized by its own coherent structure; the behavior of each of these structures determines the dynamics of an electron beam.

From Fig. 9b, we can clearly see that one of the structures (specifically, the basic structure, which was determined from the calculations of the total wavelet bicoherence) occupies a larger region in the interaction space and has a shorter time scale, $T_1 \approx 0.4$ ns. The secondary structure has a shorter spatial scale and lies between the injection plane and the basic structure. The time scale of the secondary structure, $T_2 \approx 1.0$ ns, is much longer than that of the basic structure. In this system, the coupling between these structures, whose dynamics is characterized by different time scales, is supposedly governed by the features of the complicated stochastic dynamics of an electron beam. Thus, we can say that, when the secondary structure (whose time scale $T_2$ is substantially longer than the scale $T_1$) interacts with the basic structure, it plays the role of a distributed feedback, which exerts an influence on the dynamics of an electron beam. As a result, oscillations in an electron beam become stochastic: in the case at hand, the stochastic dynamics of the beam is associated with the delayed (because of the different time scales) complicated interactions between the coherent structures forming in the system.

Now, we consider a regime arising in a system in which a high density ($n = 4.5$) ion layer lies far from the injection plane ($x_i/L = 0.25$). In this case, the Pierce parameter is equal to $\alpha = 3.0\pi$. The Fourier spectrum and the phase portrait of oscillations in the cross section $x/L = 0.25$ of the diode gap correspond to regime $C2$ (see Fig. 1 and Fig. 2h, which was obtained for the above parameter values from the oscillations of the potential in the cross section $x/L = 0.25$ of the interaction space).

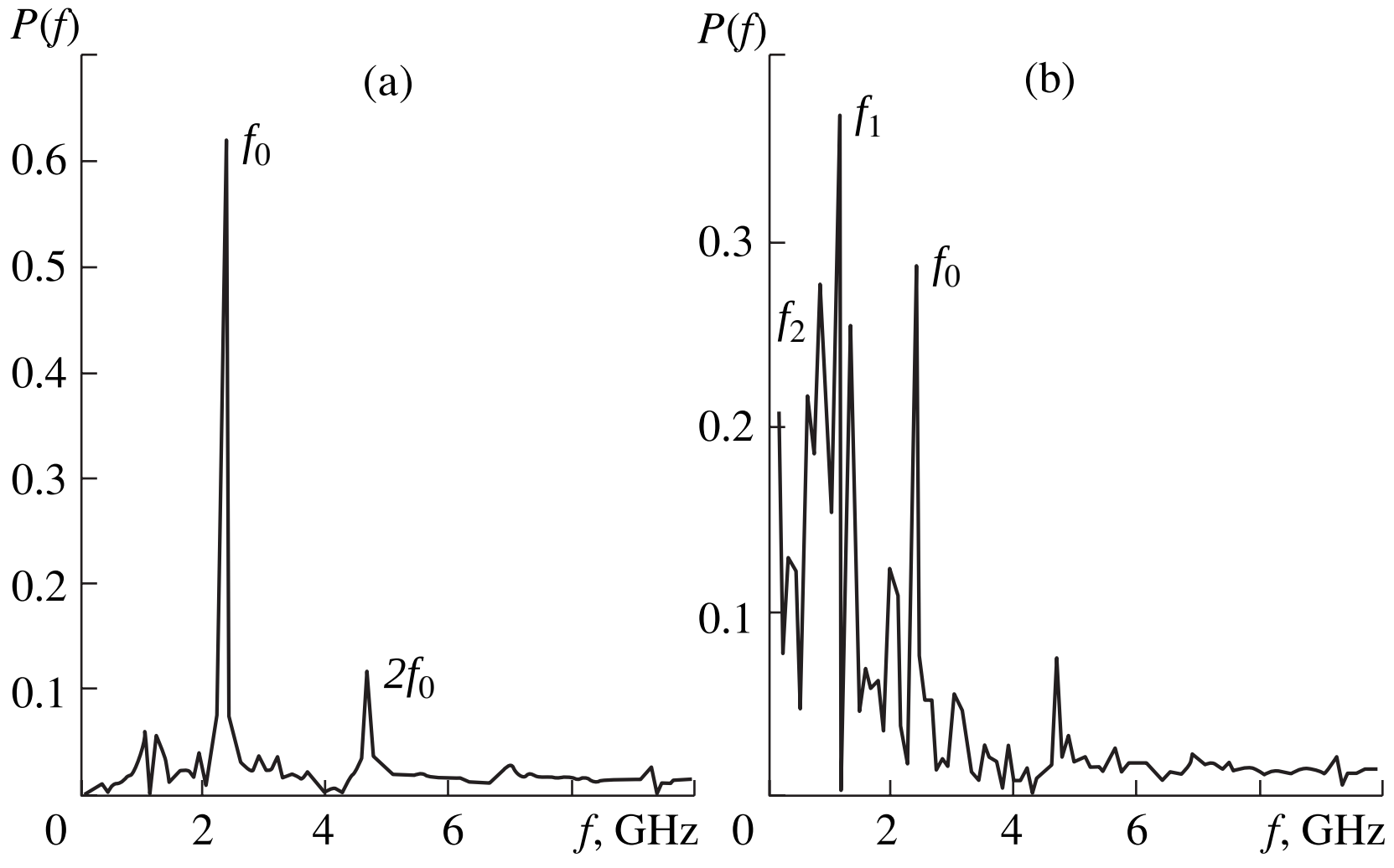

**Fig. 10.** Fourier power spectra of the space charge potential in two different cross sections of the diode gap (on a non-logarithmic scale) for $x/L$ = (a) 0.05 and (b) 0.45.

In contrast to the previous cases, the space charge field exhibits radically different temporal behavior in different cross sections of the diode gap. Thus, the spectral content of the oscillations changes both qualitatively and quantitatively along the interaction space. This is illustrated in Fig. 10, which shows the Fourier power spectra of the potential $\phi$ in the cross sections $x/L = 0.05$ (Fig. 10a) and $x/L = 0.45$ (Fig. 10b) of the diode gap. The spectra are plotted on a non-logarithmic scale. Near the injection plane, the oscillations of the potential are weakly stochastic; in the power spectrum, a sharp peak at the base generation frequency $f_0$ = 2.24 GHz in the diode gap and a lower peak at the second harmonic $2f_0$ of the base frequency are observed against a low, poorly developed, noisy pedestal. The situation changes when moving along the interaction space toward the exit grid of the system. In the power spectrum, the noisy pedestal, which decreases slowly with increasing frequency, becomes higher. In the low-frequency spectral range ($f < f_0$), spectral components appear whose energy exceeds the energy of oscillations at the frequency $f_0$, which is also present in the spectrum of oscillations of the potential in the central part of the interaction space. Thus, the amplitude of oscillations at the frequency $f_1$, predominating in the low-frequency spectral range (Fig. 10b), is larger than the oscillation amplitude at the frequency $f_0$ by a factor of 1.57.

We can suppose that such a change in the behavior of the system, namely, the generation of well-developed stochastic oscillations in the interaction space, is associated with the qualitative reconfiguration of the internal structure of an electron beam.

Let us consider the spatial dynamics of the total $b(x)$ between the oscillations of the field potential $\phi(x, t)$ and the harmonic signal. Specifically, we are interested in harmonic signals whose frequencies correspond to the characteristic frequencies in the Fourier power spectrum of the oscillations in different cross sections of the interaction space. These are the frequency $f_0$, which corresponds to the main time scale near the injection plane, and the frequency $f_1$, which determines the characteristic time scale of the dynamics of an electron beam in the central part of the interaction space.

The results of the relevant calculations are illustrated in Fig. 11. We can see that the profiles of the total wavelet bicoherence $b(x)$ calculated for harmonic signals at frequencies corresponding to different characteristic time scales of oscillations in the system are radically different.

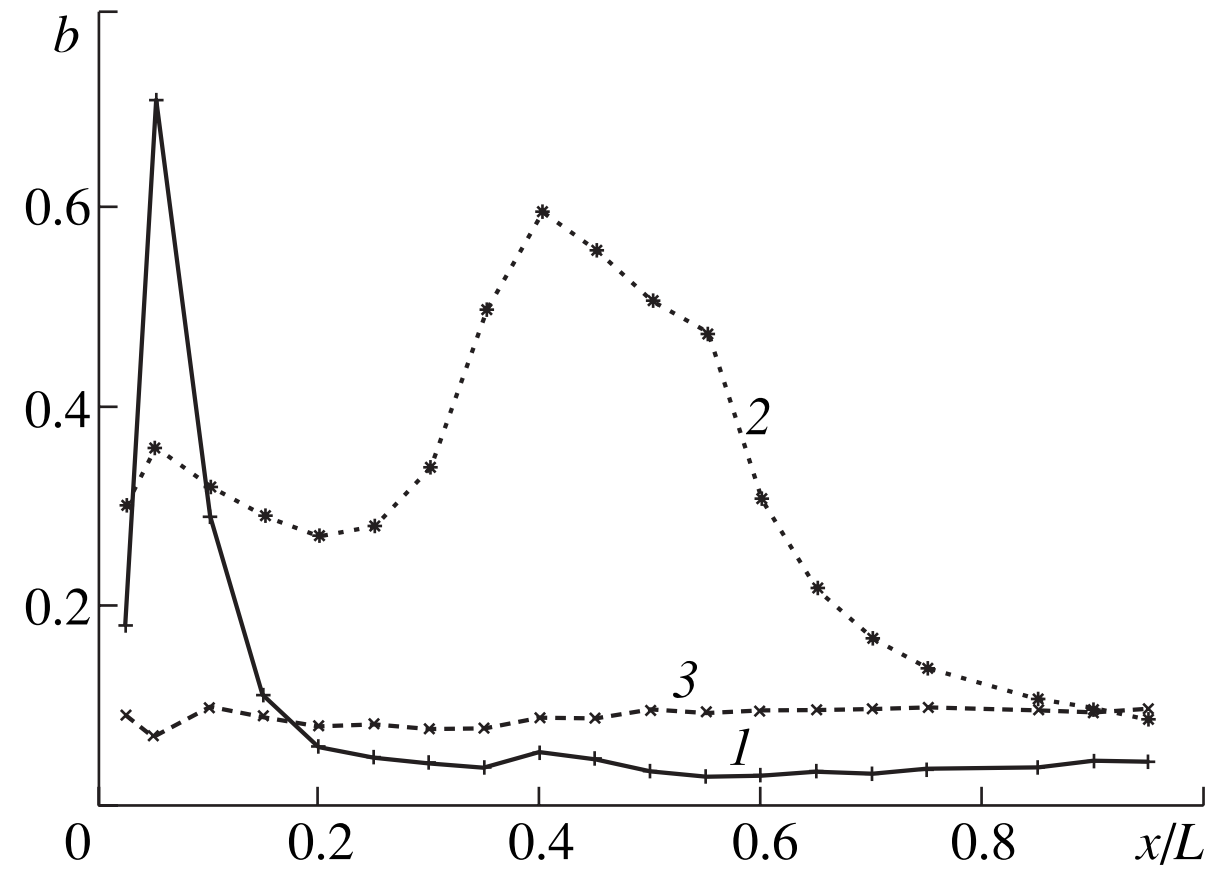

**Fig. 11.** Coordinate dependence of the total bicoherence $b(x)$ for different frequencies of the harmonic signal: (*1*) $f_0$, (*2*) $f_1$, and (*3*) $f_2$.

The profile of the total bicoherence between the oscillations of the potential and the harmonic process at the frequency $f_0$ along the interaction space (Fig. 11, curve *1*) shows that the phase coupling between the oscillations of the potential and the dynamics of the time scale $T_0 = 1/f_0$ is the strongest in the region $x/L < 0.1$. This region can be identified with the spatiotemporal structure that forms near the injection plane and whose dynamics results in the appearance of a peak in the power spectrum at the fundamental frequency $f_0$. Over the remaining part of the interaction space, the total bicoherence between the oscillations of the potential and the harmonic signal at the frequency $f_0$ is low, thereby indicating that the phase coupling between the dynamics of this coherent structure and the oscillations is weak.

The situation with the profile of the total bicoherence $b(x)$ between the oscillations of the potential and the harmonic process at the frequency $f_1$ (Fig. 11, curve *2*) is analogous to that just described. In this case, the total bicoherence is maximum in the region $x \in (0.3L, 0.6L)$. This region can be related to another coherent structure that forms in the system and whose characteristic time scale is $T_1 = 1/f_1$. Note that, in this case, the total bicoherence in the region where the first coherent structure forms is fairly large. This indicates that the second coherent structure has a fairly strong impact on the dynamics of the first structure, but, as was shown above, the first structure does not affect the second.

Hence, for the case of spatiotemporal chaos in a diode gap with a highly inhomogeneous ion background, we succeeded in resolving the coherent structures that form in the system and revealing the characteristic features of the interaction between them by exclusively analyzing the power spectrum $P(f)$ of the oscillations and the total bicoherence $b$ in different cross sections $x/L$ of the interaction space. The method proposed here for revealing the spatiotemporal coherent structures is based on the fact that the total wavelet bicoherence between the oscillations in the system and the harmonic signal at a frequency corresponding to one or another characteristic time scale of the dynamics of the system varies strongly along the interaction space.

For comparison, curve *3* in Fig. 11 shows the calculated profile of the total wavelet bicoherence between the oscillations of the potential and the harmonic signal at a frequency other than the characteristic frequencies in the power spectra that were obtained from the oscillations of the potential in different cross sections of the diode gap. To be specific, we chose the signal at the frequency marked by $f_2$ in the power spectrum shown in Fig. 10b. From Fig. 11, we can see that the bicoherence and, accordingly, the phase coupling between the oscillations in the system and the dynamics of the chosen time scale are both insignificant and vary only slightly along the interaction space. If we choose another frequency for analysis, we will be faced with a similar situation: although we will arrive at another value of bicoherence, its profile along the interaction space will be qualitatively the same, $b(x) \approx \text{const}$.

We thus can conclude that the wavelet bicoherence analysis provides an easy and convenient way of revealing the basic time scales of the dynamics of the system to be investigated. Also, in this approach, the methods of wavelet bicoherence turn out to be more advantageous than the Fourier methods for the analysis of the spatiotemporal data. The reason for this is that, in the approach proposed here, the use of the Fourier method of inferring information on coherent structures from the data sets requires a very precise determination of the corresponding characteristic time scale, which, in addition, should remain constant in time. On the other hand, in using the methods of wavelet bicoherence to determine phase coupling (15) between two oscillatory processes, there is no need to determine the characteristic time scale exactly because of the finite width of the mother wavelet (in the case at hand, this is Morlet's wavelet, see Fig. 5) in frequency space. Thus, the time scale of the process in question can vary in a fairly wide range (such that, on each of the observation scale lengths, the width of the Fourier spectrum of oscillations of the time scale is smaller than the width of the Fourier transform of the basic wavelet function).

To conclude this section, we describe the results obtained on resolving coherent structures in the spatiotemporal data. We illustrate these results by the space–time diagrams of the dynamics of an electron beam in a diode gap with an inhomogeneous ion background. To do this, we investigate the space–time diagrams of an electron beam for the cases discussed above. These diagrams are shown in Fig. 12. Each curve in the diagrams shows the trajectory of one charged particle. The concentration of the trajectories corresponds to bunches (electron structures) in an electron beam.

For the regime of regular oscillations (domain $R$ in Fig 1.), the space–time diagram of the beam (Fig. 12a) coincides qualitatively with the "classical" diagram of an electron beam in a diode gap without ion background. We can clearly see the formation of a single electron bunch—a virtual cathode—during each period of oscillations of an electron beam. The virtual cathode oscillates in both space and time and reflects most of the beam electrons back to the injection plane. The space charge density of the oscillating virtual cathode is the highest in the region $x/L \sim 0.1$, in which the majority of electrons come to a complete stop ($v \approx 0$) and are turned back. This behavior of the beam electrons agrees with the results of the wavelet bicoherence analysis, according to which coherent structure should form in the same region of the system (cf. Fig. 7).

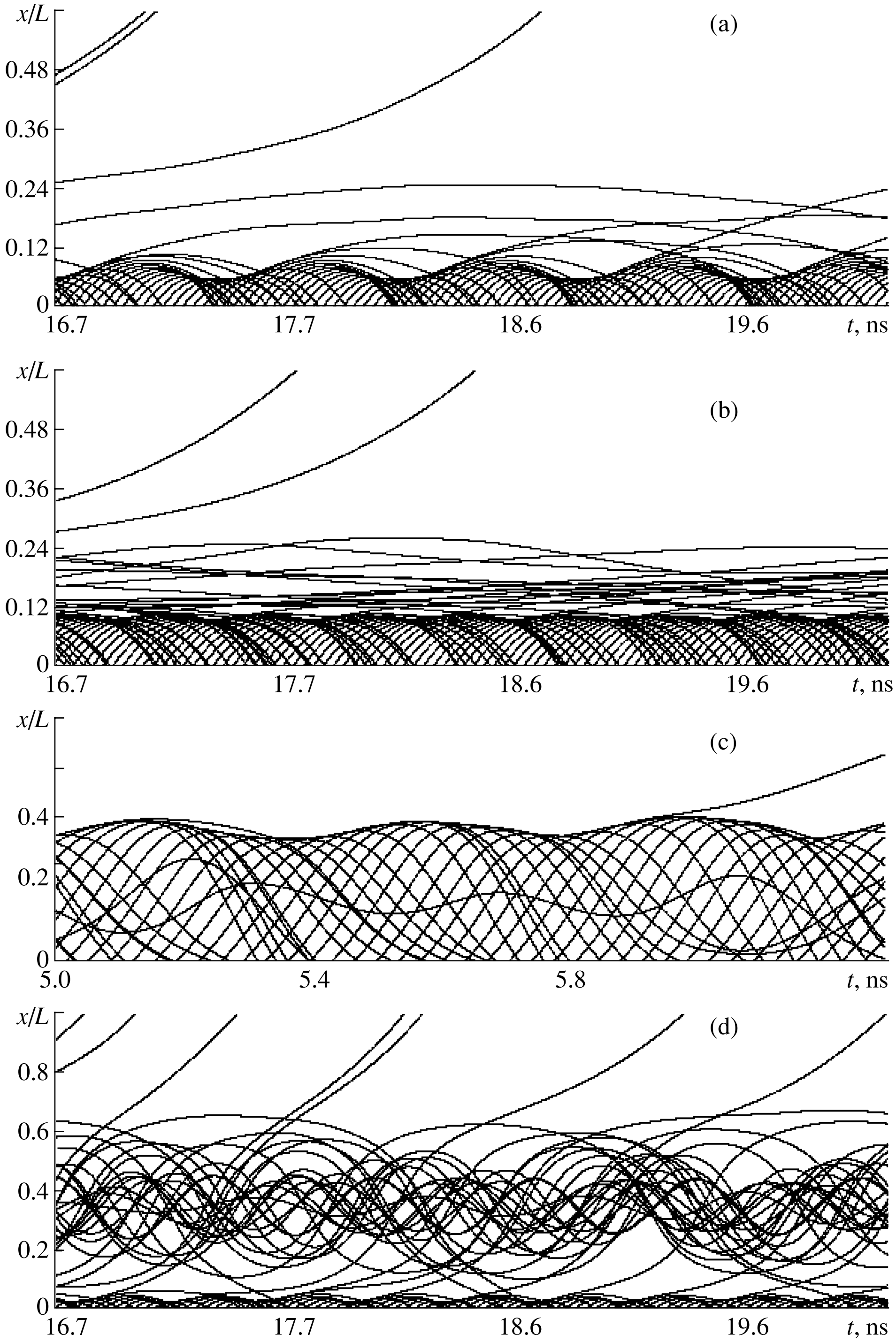

**Fig. 12.** Space–time diagrams of an electron beam in different oscillation regimes.

As the degree of inhomogeneity of the ion background increases (as the density $n$ of the ion layer becomes higher), the system passes over to regime $F$, and the dynamics of an electron beam changes. This situation is illustrated in Fig. 12b. One can see that, as before, only one electron bunch (virtual cathode) forms in the beam, but now it oscillates in space with a very small amplitude. The wavelet bicoherence analysis of the data illustrated in Fig. 8 shows that the characteristic spatial scale of the coherent structure forming in this regime is shorter than that in regime $R$. In fact, the peak of the space-charge density is closer to the injection plane; as a result, the generation frequency in the system increases sharply. Note that such a reconfiguration of the internal structure of an electron beam occurs in a jumplike manner, and the generation frequency in the diode gap also undergoes a jumplike increase (see Section 3). In this case, electron oscillations in the beam become stochastic (see the power spectrum of the oscillations and their phase portrait in Fig. 2f) because of the perturbing effect of metastable electrons that remain in the region $x/L \sim 0.2$ of the interaction space during several characteristic periods of oscillations of the main electron structure (virtual cathode). When these electrons come back to the injection plain, they start to influence the spatiotemporal dynamics of the virtual cathode in the regime at hand.

The regime of developed stochastic oscillations is a quite different situation. In regime *C3*, the density of the ion layer, which neutralizes the electron beam charge, is so high that the main structure in the beam—a virtual cathode reflecting the electrons—is "expelled" from the layer and forms at its boundary. This was clearly demonstrated by the wavelet bicoherence analysis (Fig. 9), which made it possible to resolve the basic coherent structure in an electron beam in the region $x/L \sim 0.3$. Simultaneously, a potential well with the deepest point at $x/L \approx 0.05$ forms between the virtual cathode, which lies outside the ion layer, and the injection plane. Figure 6b clearly shows that the potential well traps the electrons that have been reflected from the virtual cathode and approach the injection plane at low velocities. The trapped electrons, bouncing in the potential well, are readily seen in the space–time diagram displayed in Fig. 12c. The spatiotemporal dynamics in this regime is analogous to the dynamics of an electron beam in an oscillator based on a triode-like virtual cathode (see, e.g., [43]), in which case the potential profile is also double-humped. In a triode with a virtual cathode, the electrons trapped in the potential well form a self-similar vortex structure. Because of the interaction between the virtual cathode and the vortex, the electron dynamics in the beam becomes more complicated, by analogy with the situation described in [43]. This vortex structure can be identified with the second coherent structure, which was captured by processing the spatiotemporal data with the help of the wavelet bicoherence analysis. This interpretation is supported by a close correlation between the positions of these two structures and between the characteristic spatial and time scales of their dynamic behavior.

In regime *C2* (when the ion layer is located far from the injection plane), the electron beam dynamics is far more complicated. From Fig. 12d, we can see the formation of two electron bunches in the diode gap: a virtual cathode near the injection plane and a bunch of electrons oscillating in the potential well associated with the presence of a neutralizing ion layer with the density $n = 4.5$ in the middle of the interaction space. Comparing the space–time diagram in this figure with Fig. 11, which shows the results of a calculation of the total bicoherence for different characteristic time scales of oscillations in the system, we can conclude that the bicoherence analysis makes it possible to resolve both of the characteristic coherent structures that form in the beam.

## 6. CONCLUSION

We have investigated the nonlinear dynamics of an electron beam with an overcritical current in an inhomogeneous ion background. Diodes with an inhomogeneous distribution of the positive neutralizing background are of practical interest because they may serve as one of the simplest models of a vircator with a plasma anode [44]—a promising device of high-power electronics.

Our numerical modeling shows that such a diode exhibits different oscillatory regimes, in particular, the regime of developed spatiotemporal chaos. We have found that, as the space charge density of the ion background increases above a certain critical value, which depends on the geometry of the problem and on the electron beam current, the generation frequency in the system increases abruptly. This effect is of particular interest, because it provides the possibility of increasing the generation frequency of a device with a virtual cathode without increasing the beam current.

Applying the wavelet bicoherence analysis to a set of spatiotemporal data, we have resolved the coherent structures that form in the distributed nonlinear system under consideration, which is shown to exhibit different regimes of complicated oscillatory dynamics, and have estimated the characteristic time scales of the revealed structures and their spatial localization. The results obtained with the help of the wavelet bicoherence analysis agree well with the physical picture of the internal dynamics of an electron beam with an overcritical current in a diode gap. Thus, we have managed to identify the coherent structures that form in the system with the electron bunches in the beam.

In summary, we can say with certainty that the wavelet bicoherence analysis is an effective tool for resolving the coherent structures that govern the dynamics of spatially distributed electron-plasma systems in a state of developed chaos with electron turbulence.

## ACKNOWLEDGMENTS

This work was supported by the Russian Foundation for Basic Research, project nos. 01-02-17392 and 00-15-96673.